\documentclass[superscriptaddress,twocolumn,showpacs,pra,longbibliography]{revtex4-1}
\usepackage{bm} 
\usepackage{graphicx} 
\usepackage{amsmath}

\usepackage{amsthm}
\usepackage{amssymb} 
\usepackage{comment}
\usepackage{amsfonts}
\usepackage{dsfont}
\usepackage{color} 
\usepackage{mathtools}
\usepackage{hyperref}
\usepackage{extarrows}
\usepackage{MnSymbol}
\hypersetup{
    colorlinks=true,       
    linkcolor=cyan,          
    citecolor=magenta,        
    filecolor=magenta,      
    urlcolor=cyan,           
    runcolor=cyan
}
\usepackage{epigraph}

\newcommand {\nn}{\nonumber}

\usepackage{soul}

\begin{document}

\title{Physics-Constrained Compressed Sensing for Quantum Sensing in the Data-Starved Regime}

\author{Amir Kalev*}
\affiliation{Information Sciences Institute, University of Southern California, Arlington, VA 22203, USA}
\affiliation{Department of Physics and Astronomy, and Center for Quantum Information Science \& Technology, University of Southern California, Los Angeles, California 90089, USA}
\email[Corresponding Author:~]{amirk@isi.edu}

\begin{abstract}
\noindent Quantum sensors promise measurement sensitivities that can scale at the Heisenberg limit, but in practice their performance is often degraded by noise, finite sampling, and implementation imperfections. In this work we present a general framework for improving parameter estimation in such settings by exploiting intrinsic structural constraints of time-domain correlation functions. Our approach builds on the observation of Kemper et al. [PRL 132, 160403 (2024)] that two-time correlation functions of Hermitian observables generate Gram matrices that are positive semidefinite, a property that can be violated in experimentally acquired data. We formulate signal reconstruction as a convex optimization problem that enforces positive semidefiniteness, Toeplitz structure, and low-rank priors motivated by the underlying dynamics. We show analytically that, under suitable conditions, the ground-truth signal can be uniquely identified in the noiseless case and recovered stably in the presence of noise.  We further demonstrate numerically, in a GHZ-based magnetometry protocol, that enforcing these physical constraints can significantly improve frequency estimation from sparse and noisy data. In particular, we observe a clear advantage in the data-starved regime, where only a small number of time samples are available and standard spectral estimation methods, including matrix pencil techniques, provide limited or unstable improvement over direct fitting. While the reconstructed signals do not in general reach the shot-noise-limited performance, the proposed approach consistently reduces estimation error and recovers much of the underlying structure of the signal. These results indicate that incorporating universal physical constraints into data analysis can enhance the practical performance of quantum sensing protocols without requiring additional hardware resources or calibration.
\end{abstract}

\maketitle

\section{Introduction}
Quantum sensing uniquely leverages quantum resources to estimate physical parameters with precision surpassing classical limits. A central objective in this field is to approach the Heisenberg limit (HL), where the estimation precision scales inversely with the available quantum resources~\cite{Giovannetti2004,Fiderer2024,Giovannetti2006,Lee2002,Taylor2008,Degen2017}. In paradigmatic models like phase estimation using GHZ states, the HL represents the ultimate sensitivity achievable by fundamental quantum measurement strategy~\cite{Giovannetti2004,Leibfried2004,Giovannetti2006}. However, in practice, this limit is notoriously fragile: even modest levels of decoherence, experimental noise, or limited sampling can drastically degrade performance, often pushing sensitivity back to the Standard Quantum Limit (SQL), where the estimation precision scales inversely with the available quantum resources~\cite{Huelga1997}.

This challenge has motivated in recent years researchers to develop a broad range of quantum and classical error mitigation strategies. Quantum error correction codes tailored for metrology have been proposed to protect entangled probes against dephasing, though these schemes often require long-lived ancillas and fast feedback, which remain technologically demanding~\cite{Demkowicz2017}. Variational sensing approaches have also been devised to adaptively optimize probe states or measurement protocols to compensate for known imperfections~\cite{MacLellan2024}. Other methods recently proposed employ Bayesian estimators with prior models of the noise to reduce the mean squared error in the presence of decoherence or finite sampling noise~\cite{Ferrie2018}. While these approaches offer important advances, many rely on substantial experimental overhead or strong prior knowledge of the system and its environment~\cite{Young2012,Norris2016,Degen2017}. In practice, realistic quantum sensors are often subject to complex or time-varying noise processes that cannot be fully characterized or modeled in advance, creating a need for methods that are robust, efficient, and minimally reliant on detailed calibration. In particular, we are interested in the data-starved regime, where only a limited number of time samples are available and standard estimation methods can become unreliable.

In this work, we develop a physics-guided framework for mitigating noise in the quantum sensing framework, by exploiting fundamental structural constraints of quantum correlation functions. Rather than relying on extensive calibration or machine-learned priors, our method leverages the intrinsic positive semi-definite (PSD) nature of two-time correlation functions associated with Hermitian observables~\cite{Kemper24}. As we show, this property, which follows directly from the unitary and the inner-product structure of quantum mechanics, provides a robust and general constraint on physically realizable time-domain data. By projecting noisy experimental signals onto the space of PSD-consistent Gramian matrices, we effectively filter out unphysical distortions while preserving coherent features essential for parameter estimation. In doing so, we demonstrate that enforcing physical consistency can substantially improve the quality of reconstructed signals and the resulting parameter estimates, particularly in regimes where noise and limited sampling obscure the underlying dynamics.

Two-time correlation functions of the form 
$G(t)=\langle O(t)O(0)\rangle$
 appear broadly across quantum sensing protocols. In Ramsey interferometry, for example, the accumulated signal is directly proportional to the time-evolved expectation value of Pauli-X operator
$\langle\sigma_x(t)\rangle$, which is equivalent to a two-time correlation function when the probe is initialized in an eigenstate of 
$\sigma_x$
\cite{Lee2002,Taylor2008}. Nitrogen-vacancy-center-based magnetometry and AC field sensing often measure coherence decay or parity oscillations that are naturally expressed in terms of autocorrelation functions of spin observables~\cite{Taylor2008,Barry2020}. Dynamical decoupling spectroscopy, widely used to characterize environmental noise, recovers spectral information from time-dependent coherence signals that are the Fourier transforms of such correlators~\cite{Alvarez2011}. In all these contexts, the correlation function defines a structured Gram matrix whose physicality imposes PSD constraints (a more detailed discussion is given below). Our method capitalizes on this universal structure to denoise experimental data in a manner that is model-agnostic, calibration-free, and consistent with quantum mechanics. 

While not all quantum sensing applications reduce to correlation function estimation (e.g.,  adaptive Bayesian estimation~\cite{Huszar2012,Fiderer2024} and nonstationary control-based sensing schemes~\cite{Yang2024}), the class of protocols where time-domain signals encode parameters of interest is still significantly broad.  The formalism we propose here, applies most directly to this important subset, offering a principled and computationally efficient approach to enhance the reliability and precision of quantum sensors operating under realistic noise. To demonstrate the power of our approach, we focus on a canonical toy model: GHZ-based magnetometry~\cite{Giovannetti2004,Giovannetti2006,Lee2002,Taylor2008}. In this model,  a global phase \( \theta = \alpha t \) is imprinted on an $n$-qubit Greenberger–Horne–Zeilinger (GHZ) state via a collective spin-half Hamiltonian. In this work, we focus on scaling with respect to the number of entangled probes \(n\), which we take as the relevant quantum resource. The correlation function in time, \( G(t) = \cos(n\alpha t) \), encodes the  parameter of interest \( \alpha \), and enables its Heisenberg-limited estimation. It is known that under general noise model, the observed correlation function loses spectral resolution, and as a consequence the estimation precision can deteriorate below the HL (and depending on the noise level even below the SQL)~\cite{Huelga1997}. In this work we show that by applying compressed sensing techniques together with physical constraints, we can reconstruct a physically consistent signal \( \hat{G}(t) \) from sparse and noisy data, leading to significantly improved parameter estimation compared to direct analysis of the raw measurements.

Our findings demonstrate that enforcing minimal and general physical principles on noisy measurement data  can substantially enhance quantum sensor performance without modeling specific noise channels or requiring additional quantum resources. In addition, since our technique is based on compressed sensing, we can achieve improved estimation accuracy using substantially fewer measurements in time compared to standard approaches. Moreover, our strategy is platform-agnostic and can be applied to experimental data obtained from superconducting qubits, trapped ions, or spin ensembles. As such, it provides a practical route to enhancing the performance of quantum sensing protocols in realistic experimental settings.

\section{Theory and Methods}\label{sec:methods}

A broad class of quantum sensing protocols extract information about an unknown parameter $\phi$ by monitoring the dynamics of an observable $O(t) = e^{i H(\phi)t} O e^{-i H(\phi)t}$. The measured signal is often encoded in a two-time correlation function
\begin{equation}
    G(t) = \langle \psi_0 | O(t) O(0) | \psi_0 \rangle,
\end{equation}
where $|\psi_0\rangle$ is the initial state. Sampling at discrete times $t_k = k\Delta t$, $k=0,\dots,K-1$, yields a sequence $G_k \equiv G(t_k)$.

As noted in Ref.~\cite{Kemper24}, such correlation functions obey strong structural constraints. Defining vectors $|v_k\rangle = O(t_k)|\psi_0\rangle$, we can write

\begin{equation}\label{eq:gram_def}
    \mathcal{G}_{ij} = \langle v_i| v_j \rangle=\langle v_{i-j}| \psi_0 \rangle=G_{i-j},
\end{equation}
so that $\mathcal{G}$ is a Hermitian positive semidefinite (PSD) Gram matrix, $\mathcal{G} \succeq 0$. Moreover, stationarity implies that $\mathcal{G}$ is Toeplitz (this follows form. Thus, any physically valid correlation function corresponds to a matrix lying in the convex set
\begin{equation}
    \mathcal{C} = \{ \mathcal{G} : \mathcal{G} \succeq 0,\; \mathcal{G}\ \text{Toeplitz} \}.
\end{equation}

In many sensing settings, the correlation function admits a finite spectral decomposition
\begin{equation}
    G(t) = \sum_{\ell=1}^r a_\ell e^{-i \omega_\ell t},
\end{equation}
with $r \ll K$, for example when the dynamics is effectively supported on a small number of frequency components. In this case, the associated Gram matrix $\mathcal{G}_\star$ has rank at most $r$. This low-rank structure provides a natural prior that can be exploited to improve reconstruction from limited or noisy data.

In experiment, we observe noisy samples
\begin{equation}
    \tilde{G}_k = G_k + \epsilon_k,
\end{equation}
where the noise $\epsilon_k$ captures statistical fluctuations and experimental imperfections. We assume only that the noise is bounded in $\ell_2$ norm over the observed samples,
\begin{equation}
    \sum_{k \in \Omega} |\tilde{G}_k - G_k|^2 \leq \epsilon^2,
\end{equation}
for some index set $\Omega \subset \{0,\dots,K-1\}$ of observed time points.

The goal is to reconstruct a physically valid correlation function (equivalently, a PSD Gram matrix $\mathcal{G}$) that is consistent with the measurements while satisfying the structural constraints. As was shown in~\cite{Kalev2015}  positivity dramatically reduces the set of admissible solutions. In general, a small number of measured matrix elements may be consistent with many Hermitian matrices of different ranks. However, when the ground truth is both low rank and PSD, positivity can eliminate many otherwise admissible completions. This observation underlies compressed-sensing approaches to quantum-state tomography~\cite{Kalev2015} and is the key reason why physically constrained recovery becomes possible from a relatively small number of noisy correlation samples.

We formulate the recovery problem as a convex optimization program over the Gram matrix:
\begin{align}\label{eq:convex_program}
\hat{\mathcal{G}} = \arg\min_{\mathcal{G}} \quad & \mathrm{Tr}(\mathcal{G}) \\
\text{subject to} \quad 
& \sum_{k \in \Omega} |G_k - \tilde{G}_k|^2 \leq \epsilon^2, \nonumber \\
& \mathcal{G} = \mathrm{Toeplitz}(\{G_k\}), \nonumber \\
& \mathcal{G} \succeq 0. \nonumber
\end{align}
Since $\mathcal{G}$ is constrained to be PSD, minimizing the trace is equivalent to nuclear norm minimization for PSD matrices, thereby promoting low-rank solutions. The feasible set is convex, so Eq.~\eqref{eq:convex_program} can be solved efficiently using standard semidefinite programming or projected methods.

The convex program~\eqref{eq:convex_program} exploits both the PSD and Toeplitz structure of quantum correlation functions together with the low-rank prior. The PSD projection is implemented as a semidefinite program of size $K\times K$, solvable in $\mathcal{O}(K^3)$ time using standard convex solvers, and it is independent of the number of qubits $n$. Building on techniques from low-rank matrix recovery~\cite{Gross2010,Kalev2015} and the structured setting of Ref.~\cite{Kemper24}, one can show results of the following form (see Appendix~\ref{sec:app_a}):

\medskip
\noindent
\textbf{Proposition 1 (Exact recovery, noiseless case).} 
\textit{Let $\mathcal{G}_\star \in \mathcal{C}$ be a rank-$r$ Gram matrix. Under suitable sampling conditions on $\Omega$, there exists $|\Omega| = \mathcal{O}(r \log K)$ (up to constants depending on the sampling geometry) such that $\mathcal{G}_\star$ is the unique minimizer of Eq.~\eqref{eq:convex_program} with $\epsilon=0$.}

Here, ``suitable sampling conditions'' means that the sampled time points contain sufficient information to distinguish different low-rank PSD Toeplitz correlation functions; in particular, the sampling window must be long enough to resolve the relevant frequency content and the sampling pattern must avoid aliasing or other degeneracies.

\medskip
\noindent
\textbf{Proposition 2 (Stability, noisy case).} 
\textit{Under the same conditions, if the measurements obey $\ell_2$ noise level $\epsilon$, the solution $\hat{\mathcal{G}}$ satisfies}
\begin{equation}
    \|\hat{\mathcal{G}} - \mathcal{G}_\star\|_F \leq C \epsilon,
\end{equation}
\textit{for a constant $C$ depending on the sampling geometry.}

\medskip

These results formalize the intuition that physically constrained, low-rank correlation functions can be reconstructed from incomplete and noisy data using convex optimization, provided that the underlying assumptions on rank and sampling are satisfied.

The reconstructed correlation function $\hat{G}(t)$ can be used to estimate the parameter $\phi$. Since the reconstruction error is controlled in Frobenius norm, smooth functionals of $G(t)$, including frequency estimates and derivatives entering the Fisher information, are perturbed in a controlled manner. In particular, under regularity assumptions on $G(t)$, the induced error in estimation variance can be bounded in terms of $\epsilon$, with scaling that is at most quadratic in the noise level.

We emphasize that this approach does not increase the intrinsic quantum Fisher information of the sensing protocol. Instead, it leverages structural physical constraints to recover information that would otherwise be obscured by noise or limited sampling, thereby improving estimator performance within the assumed model class, particularly in the data-starved regime.

Following Ref.~\cite{Kemper24}, the convex program can also be used iteratively to extend the reconstructed correlation function beyond the measured time window. At each step, additional time points are inferred by enforcing consistency with the PSD and Toeplitz constraints. The accuracy and stability of this extension depend on the validity of the low-rank model, the sampling pattern, and the quality of the initial reconstruction.

In the Results section, we demonstrate how this procedure enhances spectral resolution and robustness in representative quantum sensing scenarios.

\section{Results}
To showcase the proposed method, we focus on a paradigmatic model in quantum metrology: a system of \( n \) qubits prepared in a maximally entangled generalized GHZ state and subjected to a collective phase rotation~\cite{NielsenChuang}. The dynamics of the system are governed by the Hamiltonian
\begin{align}
H = \frac{\alpha}{2} \sum_{j=1}^n Z_j,
\end{align}
where \( Z_j \) denotes the Pauli-\( Z \) operator on the \( j \)-th qubit. Under this Hamiltonian, the system evolves for a time \( t \), generating the unitary transformation \(U(t)=\exp(-i H t)\). Equivalently, one may write \(U(\theta)=\exp(-i \theta \sum_j Z_j / 2)\), where \( \theta = \alpha t \) and $\alpha$ is the unknown parameter to be estimated.  

\begin{figure*}[!th]
  \centering
\includegraphics[width=0.85\textwidth]{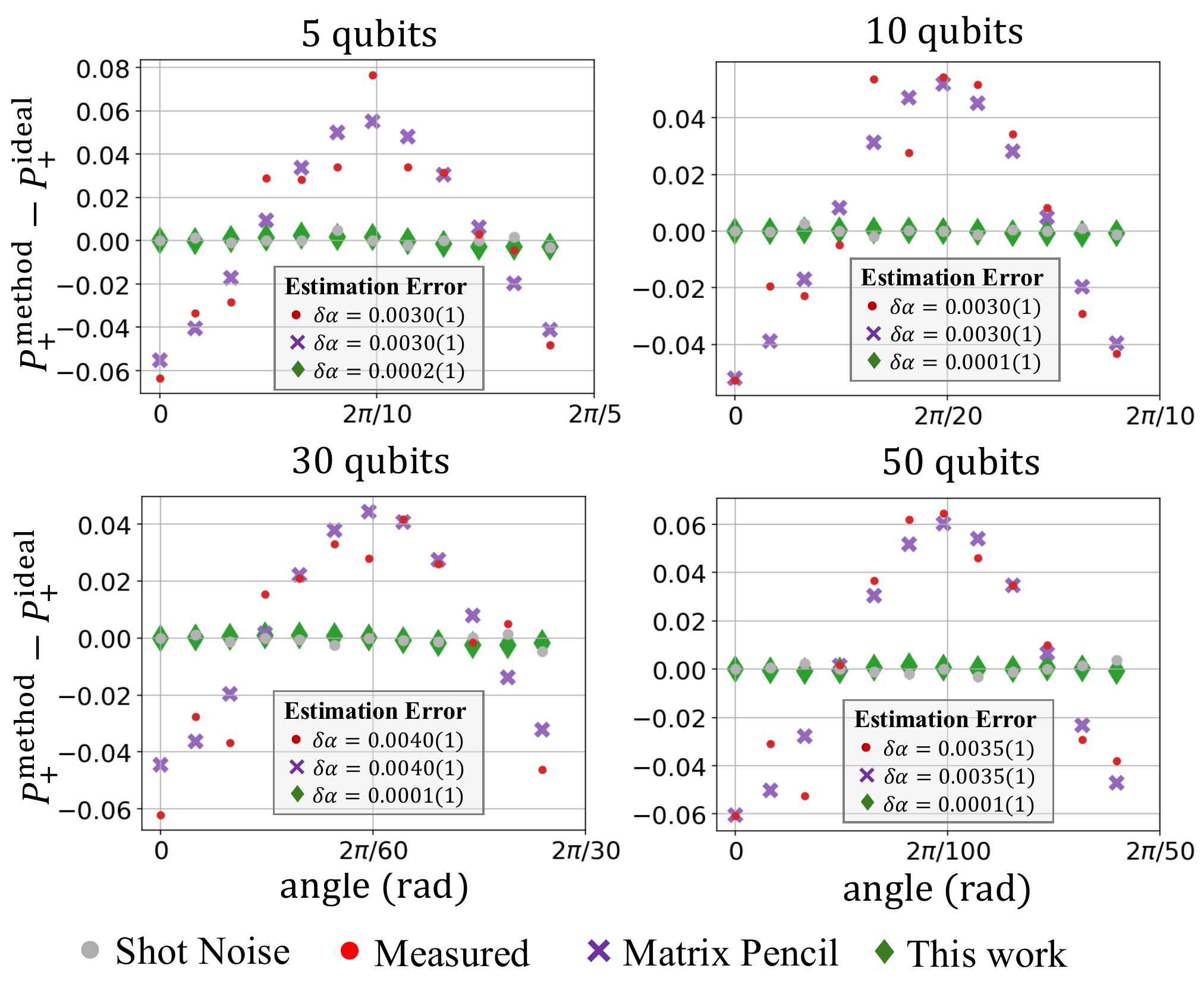}
\caption{{\bf Compressed sensing recovery of a quantum sensing signal.} Each panel shows the  difference $P_{+}^{\rm method}-P_{+}^{\rm ideal}$ in the range $[0,2\pi/n]$ for $n=5,10,30,50$ qubits. Here $P_{+}$ corresponds to the observed probability of finding the system in the $+1$  eigenspace of  $O=X^{\otimes n}$ [e.g., each shot noise point on the graph (gray dots) corresponds to $\frac1{2}(1+G^{\rm shot}_k)$, and similarly for the other data points]. This range corresponds to a single oscillation of the ideal signal, indicating quantum sensitivity that scales as $1/n$. The noisy signal (red dots) exhibits significant degradation, which is only partially addressed by standard spectral estimation (matrix pencil), but is largely corrected by the physically-constrained reconstruction (green diamonds). We observe qualitatively similar behavior across all tested values of $n$, indicating that the reconstruction procedure remains effective over a broad range of signal frequencies.}
  \label{fig:CSrecovery}
\end{figure*}
In this problem, the initial state is chosen to be the $n$-qubit GHZ state
\begin{align}
|\text{GHZ}_n\rangle = \frac{1}{\sqrt{2}}\left( |0\rangle^{\otimes n} + |1\rangle^{\otimes n} \right).
\end{align}
The measured observable is the  Pauli-\( X \) string \( O = X^{\otimes n} \), under which the GHZ state is an eigenstate with eigenvalue \( +1 \). The expectation value of \( O \) at time $t$ is given by
\begin{align}
R(\theta) = \langle \text{GHZ}_n | U^\dagger(\theta) O U(\theta) | \text{GHZ}_n \rangle = \cos(n\theta),
\end{align}
which oscillates at a frequency proportional to \( n\alpha \). The presence of \( n \) in the argument reflects the Heisenberg scaling, that is, where the sensitivity of the response function $R(\theta)$ to changes in \( \alpha \) increases linearly with the number of entangled qubits.

Since the GHZ state is a $+1$ eigenstate of the measured observable,  $R(\theta)$ can equivalently be expressed in terms of a two-time correlation function in the Heisenberg picture,
\begin{align}
G(t) = \langle \text{GHZ}_n | O(t) O(0) | \text{GHZ}_n \rangle,
\end{align}
where \( O(t) = U^\dagger(t) O U(t) \), so that $G(t)=\cos(n\alpha t)$.

In the ideal (noise-free) setting, the oscillation frequency of $G(t)=\cos(n\alpha t)$ scales linearly with $n$, so frequency-based estimators can in principle achieve a sensitivity scaling as ${\cal O}(n^{-1})$~\cite{Giovannetti2004,Giovannetti2006}. In practice, however, the measured correlation function \( \tilde{G}(t) \) is contaminated by noise arising from, e.g., decoherence, system imperfections, or control errors. These effects degrade the visibility of the signal and obscure this scaling in direct estimation from raw data, often leading to a substantial loss of spectral resolution~\cite{Huelga1997}. 

In our numerical simulations, the ideal signal was modeled as $G_k=\cos(n\alpha t_k)$, where $\{t_k\}$ are $M=50$ equally spaced points in the interval $[0,2\pi/n]$. We considered system sizes $n\in\{5,10,30,50\}$ and performed $25$ independent noise realizations for each value of $n$. To emulate limited data acquisition, only every fourth sample was retained, yielding $K=12$ input points for the reconstruction algorithm in the representative example. In addition, the data were corrupted by two noise mechanisms: (i) finite-statistics fluctuations modeled as Gaussian noise with variance $\sigma_k^2 = \frac{G_k^2(1-G_k^2)}{2500}$, and (ii) an additional additive Gaussian component with weight $\gamma=0.1$ and standard deviation $0.25$ to mimic experimental imperfections. The reconstruction was performed using the semidefinite program~\eqref{eq:convex_program} with tolerance parameter $\epsilon=0.25$. We note that the total sampling window $T$ was chosen to include approximately one full oscillation of the underlying signal for visualization. However, the PSD-based reconstruction does not rely on prior knowledge of $\alpha$ and it remains applicable for different sampling windows (as long as aliasing is not present), see App.~\ref{sec:app_b}. In addition, as a baseline for comparison, we  apply a matrix pencil method~\cite{HuaSarkar1990}. The matrix pencil is a widely-used spectral estimation technique for reconstructing sums of exponentials from sparse time samples. In our setting, this method extracts an effective oscillation frequency $\omega$ from the same $K=12$ noisy samples, without imposing physical constraints such as positive semidefiniteness or Toeplitz structure. The simulation code used to generate these results is available at~\cite{KalevGithub}.
  
To evaluate the performance of the proposed reconstruction protocol, we pass a small subset of the noisy signal $\{G_{4k}^{\rm noisy}\}_k$ for $k=[0,K-1]$ with $K=12$ through the semidefinite program~\eqref{eq:convex_program} with $\epsilon=0.25$, obtaining a solution which we denote by $\{G_{4k}^{\rm phys}\}_k$. The PSD projection is implemented as a semidefinite program of size $K\times K$, solvable in $\mathcal{O}(K^3)$ time using standard convex solvers. The value of $\epsilon$ was chosen to ensure feasibility of~\eqref{eq:convex_program} given the noise level in the simulated data. Varying it moderately, as shown in App.~\ref{sec:app_b}, yields negligible change in the variance of the estimation. We quantify performance by estimating the effective oscillation frequency $\omega$ from three different inputs: (i) direct fitting to the noisy samples, (ii) the matrix pencil estimator applied to the same noisy samples, and (iii) fitting to the PSD-constrained reconstructed signal. Since the ideal relation is $\omega=n\alpha$, errors in $\omega$ translate directly into errors in estimating $\alpha$ when $n$ is known.

Figure~\ref{fig:CSrecovery} provides a representative example of the reconstruction quality obtained from physically constrained recovery. The residual representation shows that the PSD-constrained reconstruction remains closer to the ideal correlation function than both direct fitting to noisy data and the matrix-pencil estimator in this representative instance.

\begin{figure}[t!]
  \centering
\includegraphics[width=0.95\columnwidth]{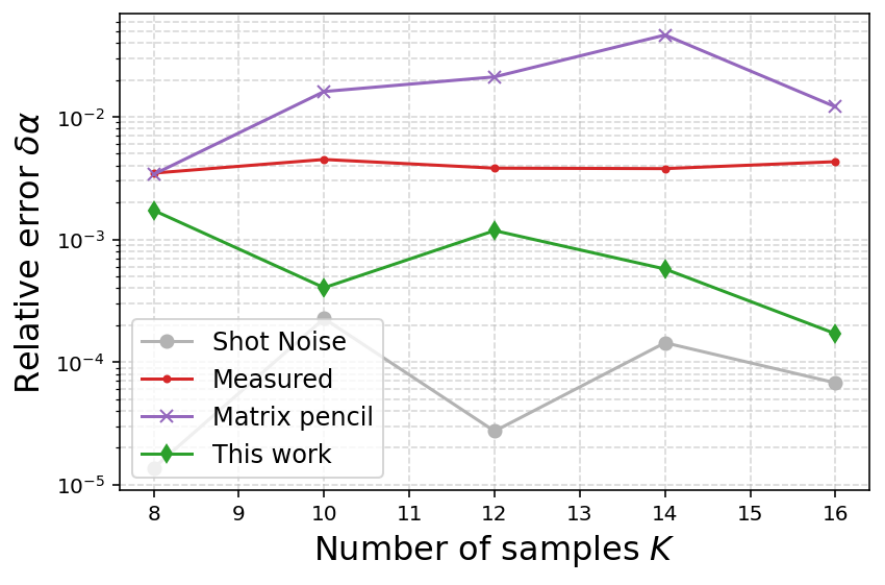}
\caption{{\bf Estimation error as a function of the number of samples.} Relative estimation error $\delta\alpha$ is shown as a function of the number of retained time samples $K$, averaged over $25$ noise realizations. In the data-starved regime ($K \sim 8$--$16$), the PSD-constrained reconstruction (green) consistently outperforms both direct fitting to noisy data (red) and the matrix-pencil estimator (purple). As $K$ increases, the performance of all methods converges, as expected when sufficient data are available.}
  \label{fig:few_data}
\end{figure}
To quantify this behavior, we examine the estimation error as a function of the number of available samples $K$, as shown in Fig.~\ref{fig:few_data}. Across all tested system sizes $n=5,10,30,50$, direct fitting to noisy data (red dots in the figure) yields errors on the order of $\delta\alpha \sim 10^{-2}$, while the matrix-pencil estimator (purple crosses) exhibits comparable or larger errors, together with noticeable variability across sampling regimes.

In contrast, the PSD-constrained reconstruction (green diamonds) provides a consistent reduction in estimation error throughout the data-starved regime. For $K$ in the range $8$--$16$, we observe improvements ranging from a factor of a few to more than an order of magnitude relative to direct fitting, with errors reaching $\delta\alpha \sim 10^{-4}$ in the best cases. As expected, the advantage diminishes as the number of available samples increases and standard estimators become more reliable.

To further characterize the regime of applicability, we performed additional numerical tests reported in Appendix~\ref{sec:app_b}. First, we mapped the performance as a function of both the number of retained samples \(K\in[6,16]\) and the imperfection strength \(\gamma\in[0.05,0.3]\). The results are consistent with the interpretation that the PSD-constrained reconstruction is useful in sparse-sampling regimes with nonzero imperfections, but that its advantage is not uniform over all \(K\) and \(\gamma\). In particular, the reconstruction often improves over the best unconstrained baseline, while some parameter choices show comparable or worse performance, as expected when the noise is either too weak for the PSD constraint to provide a visible advantage or sufficiently strong that several PSD-consistent signals become compatible with the data. Second, we tested a hidden-dephasing model \(G_\Gamma(t)=e^{-\Gamma t}\cos(n\alpha t)\), without providing \(\Gamma\) to the estimators. In this setting the PSD-constrained reconstruction continues to reduce the estimation error relative to the unconstrained baselines over the tested dephasing range, although all methods degrade as the dephasing-induced model mismatch increases.

Taken together, these results show that enforcing physical constraints can provide a substantial advantage over unconstrained spectral-estimation techniques in the data-starved regime, particularly when the measured signal contains moderate imperfections. The improvement should be interpreted as enhanced estimator performance within a structured model class, rather than a modification of the fundamental quantum limit.

\section{Summary and Conclusions}

We have developed a physics-guided framework for improving parameter estimation in quantum sensing by enforcing the positive semidefinite and Toeplitz structure of two-time correlation functions. By combining these physical constraints with convex optimization techniques inspired by compressed sensing, we reconstruct signals that are consistent with both the measured data and the underlying quantum correlation-function structure.

Our numerical results show that this approach can reduce estimation error in the presence of noise and limited sampling. In particular, we observe improvement in the data-starved regime, where only a small number of time samples are available and standard spectral estimation methods can provide limited or unstable improvement. Additional numerical tests indicate that the advantage is most pronounced for sparse sampling with moderate imperfections, while it becomes less uniform when the data are nearly ideal or when the noise/model mismatch is sufficiently strong. The hidden-dephasing tests further show that the method can remain useful beyond the ideal undamped-cosine model, although all estimators degrade as the mismatch increases.

Importantly, the improvement arises from incorporating physical consistency constraints rather than from increasing quantum resources or assuming detailed noise models. The method does not increase the intrinsic quantum Fisher information of the sensing protocol, nor does it modify the fundamental quantum limit. Rather, it improves classical inference from finite and noisy data within a structured model class. While the reconstructed signals do not, in general, achieve shot-noise-limited performance, they can recover some of the underlying structure of the signal and lead to improved estimates.

The framework is not restricted to the pure phase-shift Hamiltonian used in the GHZ example. The PSD Gram-matrix constraint applies to any Hermitian observable evolving under unitary dynamics, including Hamiltonians of the form \(H(\theta)=\theta H_1+H_2\), provided the relevant correlation functions are well defined. The Toeplitz reduction used here requires stationarity, namely dependence only on the time difference between measurements, and the compressed-sensing advantage relies on an effectively low-rank spectral representation over the measured time window. Thus, applications to more complex settings, such as criticality-based sensing or bosonic systems with broad spectral continua, are possible in principle but require case-by-case assessment.

These results indicate that enforcing general physical constraints can improve the practical performance of quantum sensing protocols in realistic experimental settings, providing a complementary approach to existing error mitigation and estimation strategies.\vfill
\bibliography{biblio}

@misc{KalevGithub,
  author = {Kalev, Amir},
  year = {2025},
  publisher = {GitHub},
  journal = {GitHub repository},
  howpublished = {\url{https://github.com/a-kalev/CS4QS}},
  commit = {}
}

@article{Degen2017,
  title = {Quantum sensing},
  author = {Degen, C. L. and Reinhard, F. and Cappellaro, P.},
  journal = {Rev. Mod. Phys.},
  volume = {89},
  issue = {3},
  pages = {035002},
  numpages = {39},
  year = {2017},
  month = {Jul},
  publisher = {American Physical Society},
  doi = {10.1103/RevModPhys.89.035002},
  url = {https://link.aps.org/doi/10.1103/RevModPhys.89.035002}
}

@article{Giovannetti2004,
author = {Vittorio Giovannetti  and Seth Lloyd  and Lorenzo Maccone },
title = {Quantum-Enhanced Measurements: Beating the Standard Quantum Limit},
journal = {Science},
volume = {306},
number = {5700},
pages = {1330-1336},
year = {2004},
doi = {10.1126/science.1104149},
URL = {https://www.science.org/doi/abs/10.1126/science.1104149},
}

@article{Giovannetti2006,
  title = {Quantum Metrology},
  author = {Giovannetti, Vittorio and Lloyd, Seth and Maccone, Lorenzo},
  journal = {Phys. Rev. Lett.},
  volume = {96},
  issue = {1},
  pages = {010401},
  numpages = {4},
  year = {2006},
  month = {Jan},
  publisher = {American Physical Society},
  doi = {10.1103/PhysRevLett.96.010401},
  url = {https://link.aps.org/doi/10.1103/PhysRevLett.96.010401}
}

@article{Demkowicz2017,
  title = {Adaptive Quantum Metrology under General Markovian Noise},
  author = {Demkowicz-Dobrza\ifmmode \acute{n}\else \'{n}\fi{}ski, Rafa\l{} and Czajkowski, Jan and Sekatski, Pavel},
  journal = {Phys. Rev. X},
  volume = {7},
  issue = {4},
  pages = {041009},
  numpages = {15},
  year = {2017},
  month = {Oct},
  publisher = {American Physical Society},
  doi = {10.1103/PhysRevX.7.041009},
  url = {https://link.aps.org/doi/10.1103/PhysRevX.7.041009}
}

@article{Lee2002,
author = {Hwang Lee and Pieter Kok and Jonathan P. Dowling},
title = {A quantum Rosetta stone for interferometry},
journal = {Journal of Modern Optics},
volume = {49},
number = {14-15},
pages = {2325--2338},
year = {2002},
publisher = {Taylor \& Francis},
doi = {10.1080/0950034021000011536},
URL = {https://doi.org/10.1080/0950034021000011536},
}

@Article{Taylor2008,
author={Taylor, J. M.
and Cappellaro, P.
and Childress, L.
and Jiang, L.
and Budker, D.
and Hemmer, P. R.
and Yacoby, A.
and Walsworth, R.
and Lukin, M. D.},
title={High-sensitivity diamond magnetometer with nanoscale resolution},
journal={Nature Physics},
year={2008},
month={Oct},
day={01},
volume={4},
number={10},
pages={810-816},
issn={1745-2481},
doi={10.1038/nphys1075},
url={https://doi.org/10.1038/nphys1075}
}

@article{Barry2020,
  title = {Sensitivity optimization for NV-diamond magnetometry},
  author = {Barry, John F. and Schloss, Jennifer M. and Bauch, Erik and Turner, Matthew J. and Hart, Connor A. and Pham, Linh M. and Walsworth, Ronald L.},
  journal = {Rev. Mod. Phys.},
  volume = {92},
  issue = {1},
  pages = {015004},
  numpages = {68},
  year = {2020},
  month = {Mar},
  publisher = {American Physical Society},
  doi = {10.1103/RevModPhys.92.015004},
  url = {https://link.aps.org/doi/10.1103/RevModPhys.92.015004}
}

@article{Alvarez2011,
  title = {Measuring the Spectrum of Colored Noise by Dynamical Decoupling},
  author = {\'Alvarez, Gonzalo A. and Suter, Dieter},
  journal = {Phys. Rev. Lett.},
  volume = {107},
  issue = {23},
  pages = {230501},
  numpages = {5},
  year = {2011},
  month = {Nov},
  publisher = {American Physical Society},
  doi = {10.1103/PhysRevLett.107.230501},
  url = {https://link.aps.org/doi/10.1103/PhysRevLett.107.230501}
}

@article{Baldwin16,
  title = {Strictly-complete measurements for bounded-rank quantum-state tomography},
  author = {Baldwin, Charles H. and Deutsch, Ivan H. and Kalev, Amir},
  journal = {Phys. Rev. A},
  volume = {93},
  issue = {5},
  pages = {052105},
  numpages = {11},
  year = {2016},
  month = {May},
  publisher = {American Physical Society},
  doi = {10.1103/PhysRevA.93.052105},
  url = {https://link.aps.org/doi/10.1103/PhysRevA.93.052105}
}

@article{Kalev2015,
  title={Quantum tomography protocols with positivity are compressed sensing protocols},
  author={Kalev, Amir and Kosut, Robert L and Deutsch, Ivan H},
  journal={npj Quantum Information},
  volume={1},
  number={1},
  pages={1--6},
  year={2015},
  publisher={Nature Publishing Group}
}

@article{Kemper24,
  title = {Denoising and Extension of Response Functions in the Time Domain},
  author = {Kemper, Alexander F. and Yang, Chao and Gull, Emanuel},
  journal = {Phys. Rev. Lett.},
  volume = {132},
  issue = {16},
  pages = {160403},
  numpages = {7},
  year = {2024},
  month = {Apr},
  publisher = {American Physical Society},
  doi = {10.1103/PhysRevLett.132.160403},
  url = {https://link.aps.org/doi/10.1103/PhysRevLett.132.160403}
}

@article{candes2009,
author = {Candes, E. J. and Plan, Y.},
title = {Tight Oracle Inequalities for Low-Rank Matrix Recovery From a Minimal Number of Noisy Random Measurements},
year = {2011},
issue_date = {April 2011},
publisher = {IEEE Press},
volume = {57},
number = {4},
issn = {0018-9448},
url = {https://doi.org/10.1109/TIT.2011.2111771},
doi = {10.1109/TIT.2011.2111771},
journal = {IEEE Trans. Inf. Theor.},
month = apr,
pages = {2342–2359},
numpages = {18},
}

@article{Gross2010,
  title = {Quantum State Tomography via Compressed Sensing},
  author = {Gross, David and Liu, Yi-Kai and Flammia, Steven T. and Becker, Stephen and Eisert, Jens},
  journal = {Phys. Rev. Lett.},
  volume = {105},
  issue = {15},
  pages = {150401},
  numpages = {4},
  year = {2010},
  month = {Oct},
  publisher = {American Physical Society},
  doi = {10.1103/PhysRevLett.105.150401},
  url = {https://link.aps.org/doi/10.1103/PhysRevLett.105.150401}
}

@book{NielsenChuang,
  title={Quantum computation and quantum information},
  author={Nielsen, Michael A and Chuang, Isaac L},
  year={2010},
  publisher={Cambridge university press}
}

@article{Huelga1997,
  title = {Improvement of Frequency Standards with Quantum Entanglement},
  author = {Huelga, S. F. and Macchiavello, C. and Pellizzari, T. and Ekert, A. K. and Plenio, M. B. and Cirac, J. I.},
  journal = {Phys. Rev. Lett.},
  volume = {79},
  issue = {20},
  pages = {3865--3868},
  numpages = {0},
  year = {1997},
  month = {Nov},
  publisher = {American Physical Society},
  doi = {10.1103/PhysRevLett.79.3865},
  url = {https://link.aps.org/doi/10.1103/PhysRevLett.79.3865}
}

@Article{MacLellan2024,
author={MacLellan, Benjamin
and Roztocki, Piotr
and Czischek, Stefanie
and Melko, Roger G.},
title={End-to-end variational quantum sensing},
journal={npj Quantum Information},
year={2024},
month={Nov},
day={19},
volume={10},
number={1},
pages={118},
issn={2056-6387},
doi={10.1038/s41534-024-00914-w},
url={https://doi.org/10.1038/s41534-024-00914-w}
}

@article{Ferrie2018,
doi = {10.1088/1367-2630/aaf207},
url = {https://dx.doi.org/10.1088/1367-2630/aaf207},
year = {2018},
month = {dec},
publisher = {IOP Publishing},
volume = {20},
number = {12},
pages = {123005},
author = {Ferrie, Chris and Granade, Chris and Paz-Silva, Gerardo and Wiseman, Howard M.},
title = {Bayesian quantum noise spectroscopy},
journal = {New Journal of Physics},
}

@article{Young2012,
  title = {Qubits as spectrometers of dephasing noise},
  author = {Young, Kevin C. and Whaley, K. Birgitta},
  journal = {Phys. Rev. A},
  volume = {86},
  issue = {1},
  pages = {012314},
  numpages = {7},
  year = {2012},
  month = {Jul},
  publisher = {American Physical Society},
  doi = {10.1103/PhysRevA.86.012314},
  url = {https://link.aps.org/doi/10.1103/PhysRevA.86.012314}
}

@article{Norris2016,
  title = {Qubit Noise Spectroscopy for Non-Gaussian Dephasing Environments},
  author = {Norris, Leigh M. and Paz-Silva, Gerardo A. and Viola, Lorenza},
  journal = {Phys. Rev. Lett.},
  volume = {116},
  issue = {15},
  pages = {150503},
  numpages = {5},
  year = {2016},
  month = {Apr},
  publisher = {American Physical Society},
  doi = {10.1103/PhysRevLett.116.150503},
  url = {https://link.aps.org/doi/10.1103/PhysRevLett.116.150503}
}

@article{Huszar2012,
  title = {Adaptive Bayesian quantum tomography},
  author = {Husz\'ar, F. and Houlsby, N. M. T.},
  journal = {Phys. Rev. A},
  volume = {85},
  issue = {5},
  pages = {052120},
  numpages = {5},
  year = {2012},
  month = {May},
  publisher = {American Physical Society},
  doi = {10.1103/PhysRevA.85.052120},
  url = {https://link.aps.org/doi/10.1103/PhysRevA.85.052120}
}

@article{Fiderer2024,
  title = {Neural-Network Heuristics for Adaptive Bayesian Quantum Estimation},
  author = {Fiderer, Lukas J. and Schuff, Jonas and Braun, Daniel},
  journal = {PRX Quantum},
  volume = {2},
  issue = {2},
  pages = {020303},
  numpages = {15},
  year = {2021},
  month = {Apr},
  publisher = {American Physical Society},
  doi = {10.1103/PRXQuantum.2.020303},
  url = {https://link.aps.org/doi/10.1103/PRXQuantum.2.020303}
}

@Article{Yang2024,
author={Yang, Xiaodong
and Long, Xinyue
and Liu, Ran
and Tang, Kai
and Zhai, Yue
and Nie, Xinfang
and Xin, Tao
and Li, Jun
and Lu, Dawei},
title={Control-enhanced non-Markovian quantum metrology},
journal={Communications Physics},
year={2024},
month={Aug},
day={21},
volume={7},
number={1},
pages={282},
issn={2399-3650},
doi={10.1038/s42005-024-01758-8},
url={https://doi.org/10.1038/s42005-024-01758-8}
}

@article{
Leibfried2004,
author = {D. Leibfried  and M. D. Barrett  and T. Schaetz  and J. Britton  and J. Chiaverini  and W. M. Itano  and J. D. Jost  and C. Langer  and D. J. Wineland },
title = {Toward Heisenberg-Limited Spectroscopy with Multiparticle Entangled States},
journal = {Science},
volume = {304},
number = {5676},
pages = {1476-1478},
year = {2004},
doi = {10.1126/science.1097576},
URL = {https://www.science.org/doi/abs/10.1126/science.1097576},
eprint = {https://www.science.org/doi/pdf/10.1126/science.1097576},
}

@article{HuaSarkar1990,
  author={Hua, Yingbo and Sarkar, Tapan K.},
  journal={IEEE Transactions on Acoustics, Speech, and Signal Processing}, 
  title={Matrix pencil method for estimating parameters of exponentially damped/undamped sinusoids in noise}, 
  year={1990},
  volume={38},
  number={5},
  pages={814-824},
  doi={10.1109/29.56027}
}

\appendix
\section{Compressed Sensing of a PSD Matrix}\label{sec:app_a}
\subsection{Background}
\begin{figure}[!ht]
  \centering
\includegraphics[width=0.48\textwidth]{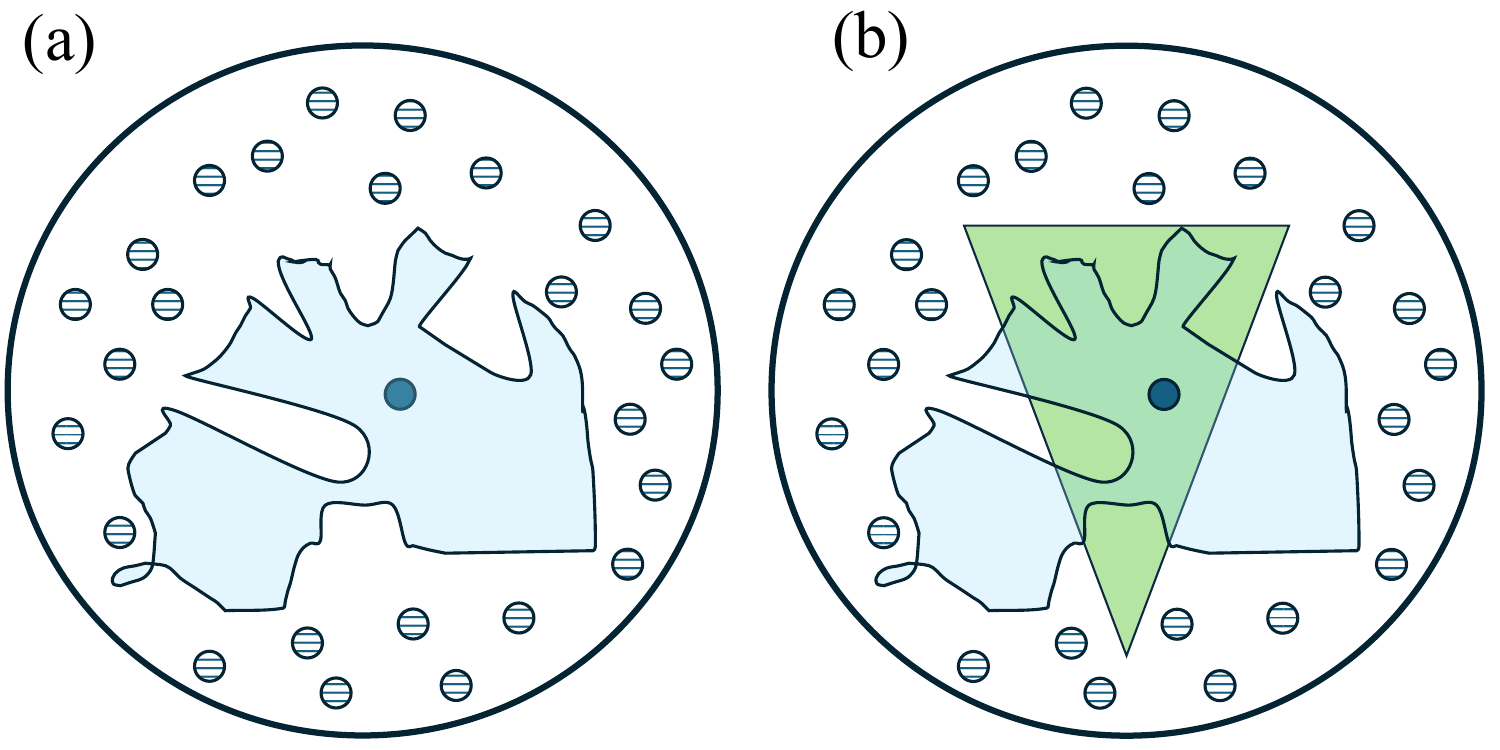}
\caption{{\bf The power of positivity (an illustration).} (a) The outer circle represents the set of $d \times d$ Hermitian matrices, the blue region represents the non-convex subset of those matrices whose rank is equal or smaller than $r$, the circle within the rank-$r$ subset represents the ground truth matrix. This matrix can be uniquely determined, within the subset of rank-$r$ matrices, by $O(rd)$ real numbers. However, there are infinitely many matrices whose rank is strictly larger than $r$ that are consistent with these numbers. These matrices are represented by the circles outside the blue region.  (b) It was shown in~\cite{Kalev2015} that if the ground truth is a PSD matrix, then $O(rd)$ real numbers uniquely determine it within the entire convex set of $d \times d$ PSD matrices, of {\it any} rank.}
  \label{fig:PoP}
\end{figure}
Compressed sensing (CS) is a signal processing technique that allows for the efficient acquisition and reconstruction of a signal by exploiting its low-rankness (in the case where the signal is represented by a Hermitian matrix)~\cite{candes2009}. Consider a $d\times d$ Hermitian matrix $\rho_\star$ whose rank is lower or equal than some $r\ll d$. When a signal is low-rank (``compressible''), CS methods~\cite{candes2009} can recover it from $O(rd)$ samples,  significantly fewer than the $O(d^2)$ parameters required to specify a generic Hermitian matrix. This is achieved by (1) designing data acquisition schemes that obey special properties, such as the Restricted Isometry Property (RIP) and (2) solving a nuclear-norm minimization program that looks for the solution that is consistent with acquired data:
\begin{align}\label{eq:cs}
\hat{\rho}&=\text{argmin}~\Vert\rho\Vert_*\\\nn
&\text{such that:} \Vert{\cal A}[\rho]-y\Vert_2\leq\epsilon,
\end{align}
where $\Vert\cdot\Vert_*$ is the nuclear-norm, ${\cal A}$ is the acquisition map ${\cal A}: {\mathbb C}^{d\times d}\to {\mathbb R}^{O(rd)}$, and $y\in {\mathbb R}^{O(rd)}$ is the acquired samples.
The CS methodology~\cite{candes2009} assures that when $\text{Rank}(\rho_\star)\leq r$ then if ${\cal A}$ satisfies RIP (or other similar property) the solution of the program~\eqref{eq:cs} is $\epsilon$-close to the ground truth in Frobenius norm, 
\begin{align}\label{eq:eps-close}
    \Vert\hat{\rho}-\rho_\star\Vert_F \leq C\,\epsilon.
\end{align}
This is schematically illustrated in Fig.~\ref{fig:PoP} (a). One of the key innovations of the CS method is the use of a convex optimization program to obtain a robust solution to a generally NP-hard problem of low-rank matrix recovery.

Applying the CS methodology to the context of quantum state tomography, where the goal is to recover a PSD matrix with a unit trace, the authors of Ref.~\cite{Kalev2015} proved that the PSD constraint in fact enables CS. There, the authors proved that under the same (rank and RIP) constraint, when in addition $\rho_\star\geq 0$ then in absence of noise $\rho_\star$ is uniquely specified within the feasible set, see Fig.~\ref{fig:PoP} (b) for an illustration. This implies that the feasibility problem
\begin{align}\label{eq:feasibility}
\text{Find}&~\rho\\\nn
\text{such that:}&~ {\cal A}[\rho]=y,\; \rho\geq0
\end{align}
has a unique solution $\hat{\rho}=\rho_\star$. This is a feasibility program defined on the entire cone of PSD matrices (implicitly in the program $\rho$ is of dimension $d$). From this follows, as was proven in~\cite{Gross2010,Kalev2015}
that the solution to
\begin{align}\label{eq:cs+psd}
\hat{\rho}&=\text{argmin Tr}(\rho)\\\nn
&\text{such that:}~ \Vert{\cal A}[\rho]-y\Vert_2\leq\epsilon,\, \rho\geq 0
\end{align}
is $\epsilon$-close to the ground truth, according to Eq.~\eqref{eq:eps-close}.

\subsection{Application to Gram matrix recovery}

This subsection follows closely the methods presented in Ref.~\cite{Baldwin16}. There, the authors  extended the application of CS to PSD matrices where the acquisition map does not follow RIP (or other related properties). Instead, the acquired data corresponds to specific matrix elements (in a given basis of representation). In this setting, the problem reduces to a matrix completion problem with the additional constraint that the ground truth is a PSD matrix.  

Consider a block representation of a $d\times d$ matrix $\rho$:
\begin{align}
    \rho=\begin{pmatrix}
        A&B\\B^\dagger&C
    \end{pmatrix},
\end{align}
assuming  $\rho$ is a Hermitian square matrix of dimension $d$ for concreteness. The Haynsworth inertia additivity formula ensures that the rank of $\rho$ satisfy the relation
\begin{align}\label{eq:Hay}
    \text{Rank}(\rho)=\text{Rank}(A)+\text{Rank}(\rho/A)
\end{align}
where $\rho/A=C-B^\dagger A^{-1}B$ is the Schur complement of $\rho$ (assuming $A$ is 
non-singular, i.e., invertible).

Consider the case where $\text{Rank}(\rho)= r$ for some known $r\ll d$. In this case, we can take $A$ to be of size $r\times r$, and then it is guaranteed that, when it is invertible, $\text{Rank}(A)=r$. Therefore, with this choice of $A$, $\text{Rank}(\rho/A)=0$ and it follows that
\begin{align}\label{eq:Schur=0}
  C=B^\dagger A^{-1}B.  
\end{align}
This equation implies that the first $r$ rows and columns of $\rho$, when $\text{Rank}(\rho)= r$, uniquely determine it within the set of rank-$r$ matrices. However, while Eq.~\eqref{eq:Schur=0} guarantees that there is no other rank-$r$ (Hermitian) matrix with the same first $r$ rows and columns as $\rho$,  there may be (infinitely many) other hermitian matrices whose rank is strictly larger than $r$ that have the same first $r$ rows and columns as $\rho$ (but different from it on the lower-right sub-matrix of dimension $d-r$, generically denoted here as $C$). In Ref.~\cite{Baldwin16} it was shown that if $\rho$ is PSD matrix of rank $r$, then its first $r$ rows and columns (or any of its $r$ rows and corresponding columns, for that matter) uniquely specify it within the set of PSD matrices of dimension $d$, regardless of their rank, under the stated assumptions. This result is rooted in the same principles of the CS result discussed earlier; see Fig.~\ref{fig:PoP} (b) for illustration. 

The implication of the above result can be summarized as follows. Consider a ground truth $d$ dimensional PSD Hermitian matrix $\rho_\star$ of rank-$r$, whose first $r$ rows and columns are know (we  cast them into $A_\star,B_\star,$ and $B_\star^\dagger$ as above). Then the results of~\cite{Baldwin16} imply that the feasibility problem 
\begin{align}\label{eq:feas}
    \text{Find}&~C\\\nn
    \text{such that:}&~\rho=\begin{pmatrix} A_\star&B_\star\\B_\star^\dagger&C\end{pmatrix},~\rho\geq0,~ (C=C^\dagger)
\end{align}
has a unique solution $\hat{C}=C_\star$ (using an obvious notation), with $\text{Rank}(\rho_\star)=r$. Note that \eqref{eq:feas} is a convex program, in fact a semi-definite program, and therefore, its feasibility set is the convex set of $d$-dimensional Hermitian PSD matrices. Hence, this unique solution exists within the set of all PSD Hermitian matrices (of dimension $d$), of any rank.

The proof of the above statement follows from the Haynsworth inertia additivity formula which states that the number of  positive/zero/negative eigenvalues of $\rho$ equals that of $A$ plus that of $\rho/A$, respectively. Since in the PSD case, by construction, the number of positive eigenvalues of $A_\star$ equals to the number of positive eigenvalues of $\rho_\star$, all other matrices of rank strictly larger than $r$, that  share with $\rho_\star$ its submatrices $A_\star$ and $B_\star$, must have negative eigenvalues, and therefore, reside outside the feasibility set. 
 
In practice we may know $A_\star$ and $B_\star$ up to some precision. Since in the absence of noise $C_\star$ can be uniquely found, e.g., by solving a convex optimization problem [essentially, of the form of~\eqref{eq:feas}], it was proven in Ref.~\cite{Baldwin16} that in the presence of noise the  solution of a properly designed convex numerical program, such as the one described in~\eqref{eq:cs+psd} (where the  map ${\cal A}$ corresponds to acquiring the matrix elements in the first row and column of $\rho_\star$), would be within $\epsilon$ (in Frobenius norm) of the ground truth $\rho_\star$.

To connect these results to the recovery of a PSD   Gram matrix ${\cal G}_\star$ we first assume its rank is not larger than some $r\ll M$, where $M$ is determined form the size of the acquired (noisy) time-series data, $\{\tilde{G}_k\}$. We then note that by construction the  ground truth Gram matrix ${\cal G}_\star$ is a Toeplitz matrix  constructed from the (noiseless) time series $\{(G_\star)_k\}$. Therefore, the acquired time series corresponds directly to the rows and columns of the corresponding Gram matrix. For simplicity, we further assume here  that ${\cal G}_\star$ is a real  matrix, since this is the prevalent case in quantum sensing. A generalization to Hermitian matrix is immediate.  

With this structure of ${\cal G}_\star$, we can directly import the results form Ref.~\cite{Baldwin16} to the case of quantum sensing.  In particular, following Ref.~\cite{Baldwin16}, for the case where $\text{Rank}({\cal G}_\star)=r$, since ${\cal G}_\star\geq 0$,   the time-series $\{(G_\star)_k\}$ with $k\in[1,r]$ determines ${\cal G}_\star$ in the noiseless case under the stated low-rank and PSD assumptions. In addition, in the noisy case, the results of Ref.~\cite{Baldwin16}  imply that we can denoise the time series signal by acquiring (at least) $r$ data points, $\tilde{G}_1,\ldots,\tilde{G}_{r}$. From our discussion above, solving 
\begin{align}\label{eq:G+psd}
\hat{{\cal G}}&=\text{argmin Tr}({\cal G})\\\nn
&\text{such that:}~ \sum_{k=1}^{r} |\tilde{G}_k-G_k|^2\leq\epsilon^2,\\\nn&\;\;\;\;\;\;\;\;\;\;\;\;\;\;\;\;\;\;\;
{\cal G}=\text{Toeplitz} (\{G_k\}),\, {\cal G}\geq 0,
\end{align}
where $\text{Toeplitz} (\{g_k\})$ is an operation that takes a series $\{g_k\}$ to a corresponding Toeplitz matrix, is guaranteed to result with a robust estimation $\hat{{\cal G}}$ such that $\Vert \hat{\cal G}-{\cal G}_\star\Vert_F \leq C\,\epsilon$. 

Finally, following the techniques developed in~\cite{Kemper24}, we can adapt the convex program~\eqref{eq:G+psd} to  robustly extend the time series (one time-step at a time) to larger times. As expected, increasing the number of acquired data points can improve the stability and extent of this extrapolation.

\section{Additional Numerical Results}\label{sec:app_b}
We include a few  numerical results that complement the results in the main text. 
\subsection{Partial-period sampling}
\begin{figure}[!th]
  \centering
\includegraphics[width=0.95\columnwidth]{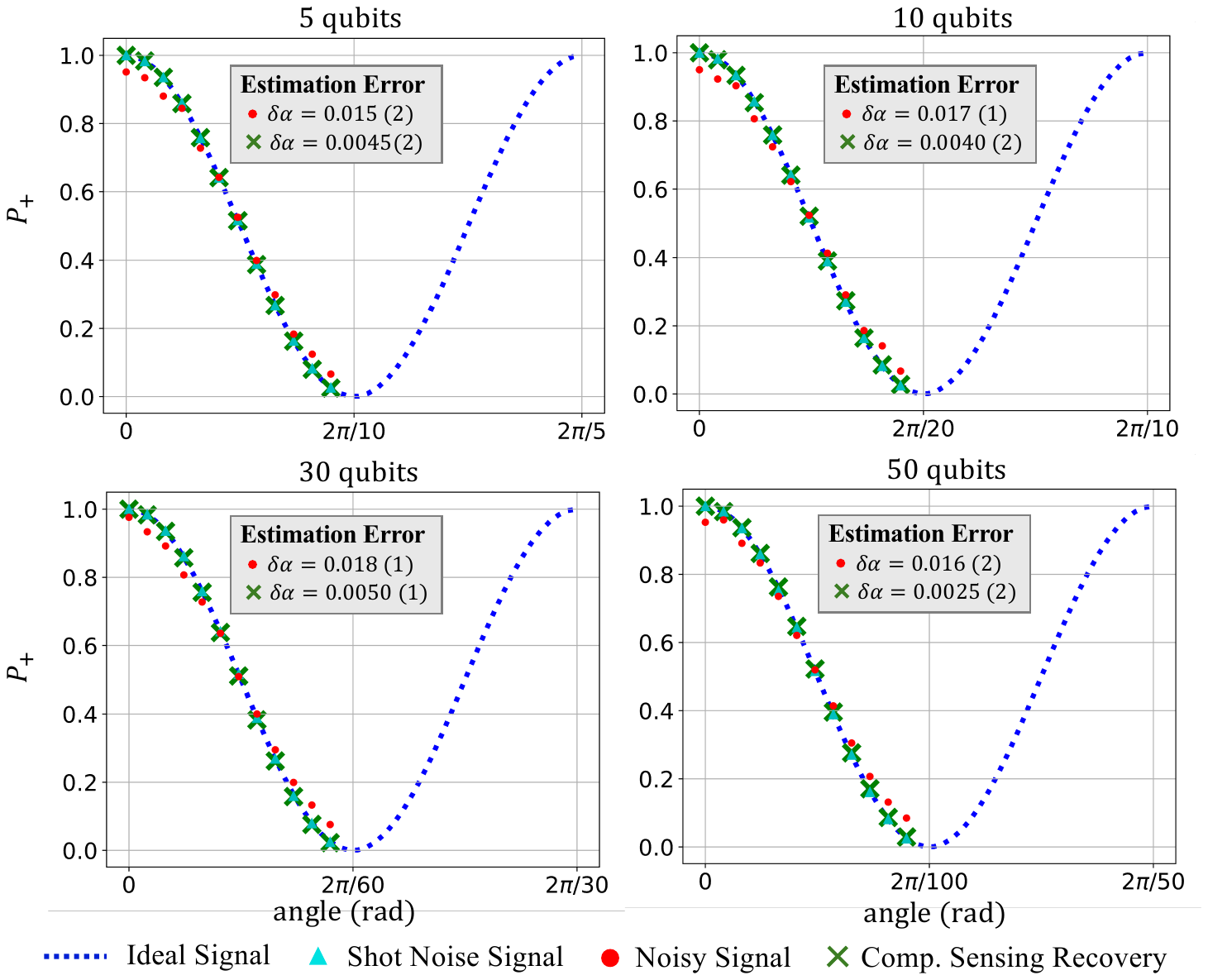}
\caption{{\bf Recovery of a quantum sensing signal.} The PSD-based protocol is shown here to be robust for cases where the observations are limited to a fraction of the oscillation period. In this numerical example it reduces the estimation variance, by approximately one order of magnitude, even in the partial-oscillation regime.}
  \label{fig:CSrecovery_app}
\end{figure}

In Fig.~\ref{fig:CSrecovery_app} we plot the variance $\delta \alpha$ when the observations are limited to a fraction of the oscillation period.  If the measurement covers only a fraction of an oscillation, the correlation function remains analytic and well-defined, and our PSD-based method, can be used to enforce global consistency of the data.  We have confirmed this behavior numerically as shown in Fig.~\ref{fig:CSrecovery_app}. The results show that our method can reduce the estimation variance, by approximately one order of magnitude, even in the partial-oscillation regime. We must note that, of course, if the sampling rate is too low (violating the Nyquist criterion), multiple frequencies can fit the data equally well (``aliasing''). Our framework does not remove this intrinsic ambiguity.
 
\subsection{Dependence on the SDP tolerance}
\begin{figure}[!th]
  \centering
\includegraphics[width=0.85\columnwidth]{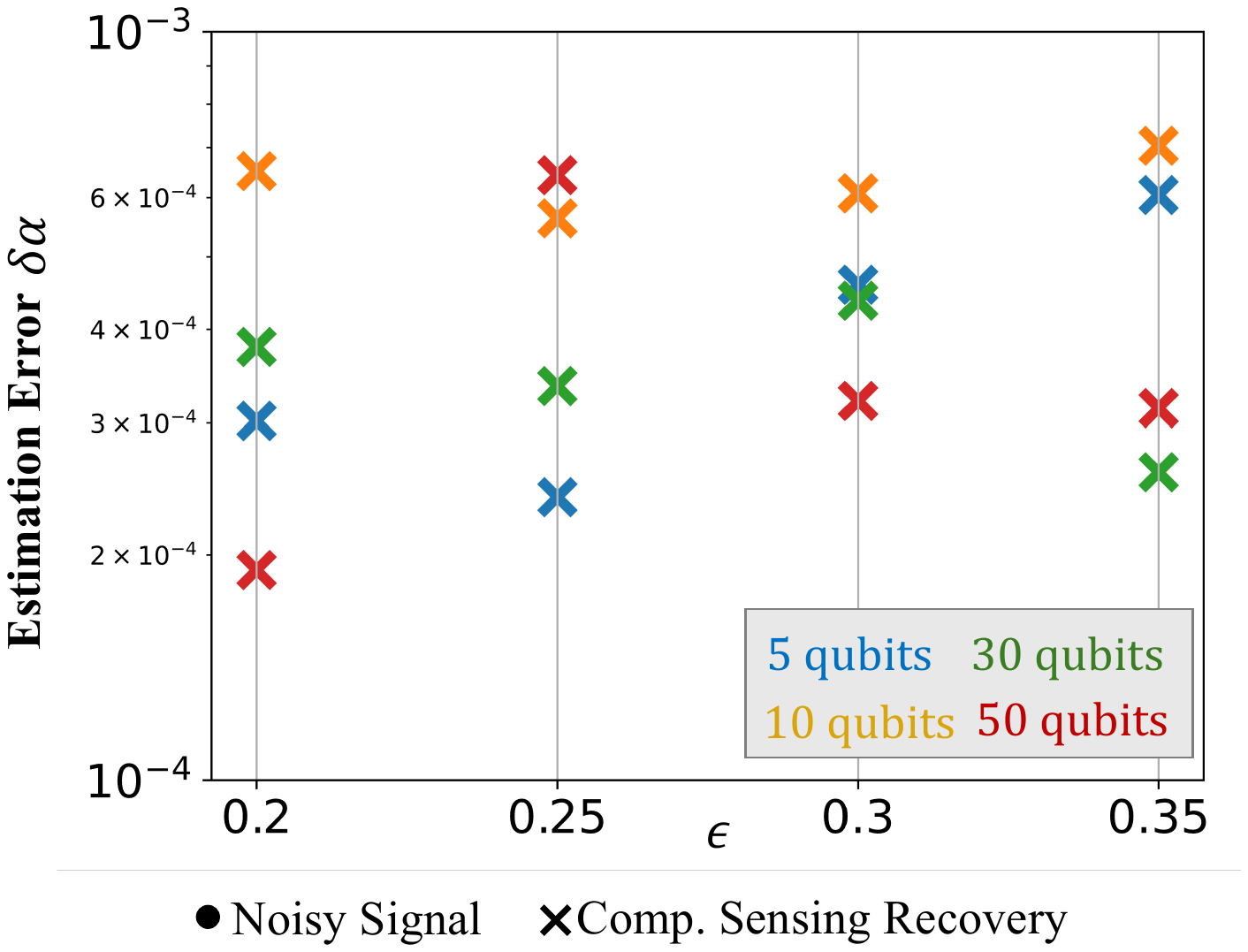}
\caption{{\bf Recovery of a quantum sensing as a function of $\epsilon$.} The variance of the estimated parameter is shown to be robust under moderate variation of $\epsilon$.}
  \label{fig:epsilon}
\end{figure}

Figure~\ref{fig:epsilon} shows the dependence of the estimation variance on the numerical  parameter $\epsilon$. The parameter \(\epsilon\) effectively determines how close the PSD solution is to the measured data, and  it is set relative to the noise variance. In our simulations in the main text we used \(\epsilon = 0.25\) which provides stable results; varying it moderately, as shown in Fig.~\ref{fig:epsilon}, yields negligible change in the variance of the estimation. The remaining one-order-of-magnitude gap to the shot-noise-limited sensitivity arises from finite sampling noise, not from the choice of \(\epsilon\). 

\subsection{Operating regime in sample number and imperfection strength}
To further characterize the regime in which the PSD-constrained reconstruction is useful, we performed a two-dimensional numerical scan over the number of retained time samples \(K\) and the imperfection-noise strength \(\gamma\). In this test, we used
\[
K\in\{6,8,10,12,14,16\},
\qquad
\gamma\in[0.05,0.3],
\]
with six equally spaced values of \(\gamma\). For each pair \((K,\gamma)\), the same \(K\) noisy samples were supplied to all estimators: direct fitting to the noisy data, the matrix-pencil estimator, and the PSD-constrained reconstruction. We quantified usefulness by comparing the PSD-constrained reconstruction $\delta\alpha_{\rm CS}$ to the best matrix-pencil baseline,
$\delta\alpha_{\rm MP}$.

\begin{figure}[!th]
  \centering
\includegraphics[width=0.95\columnwidth]{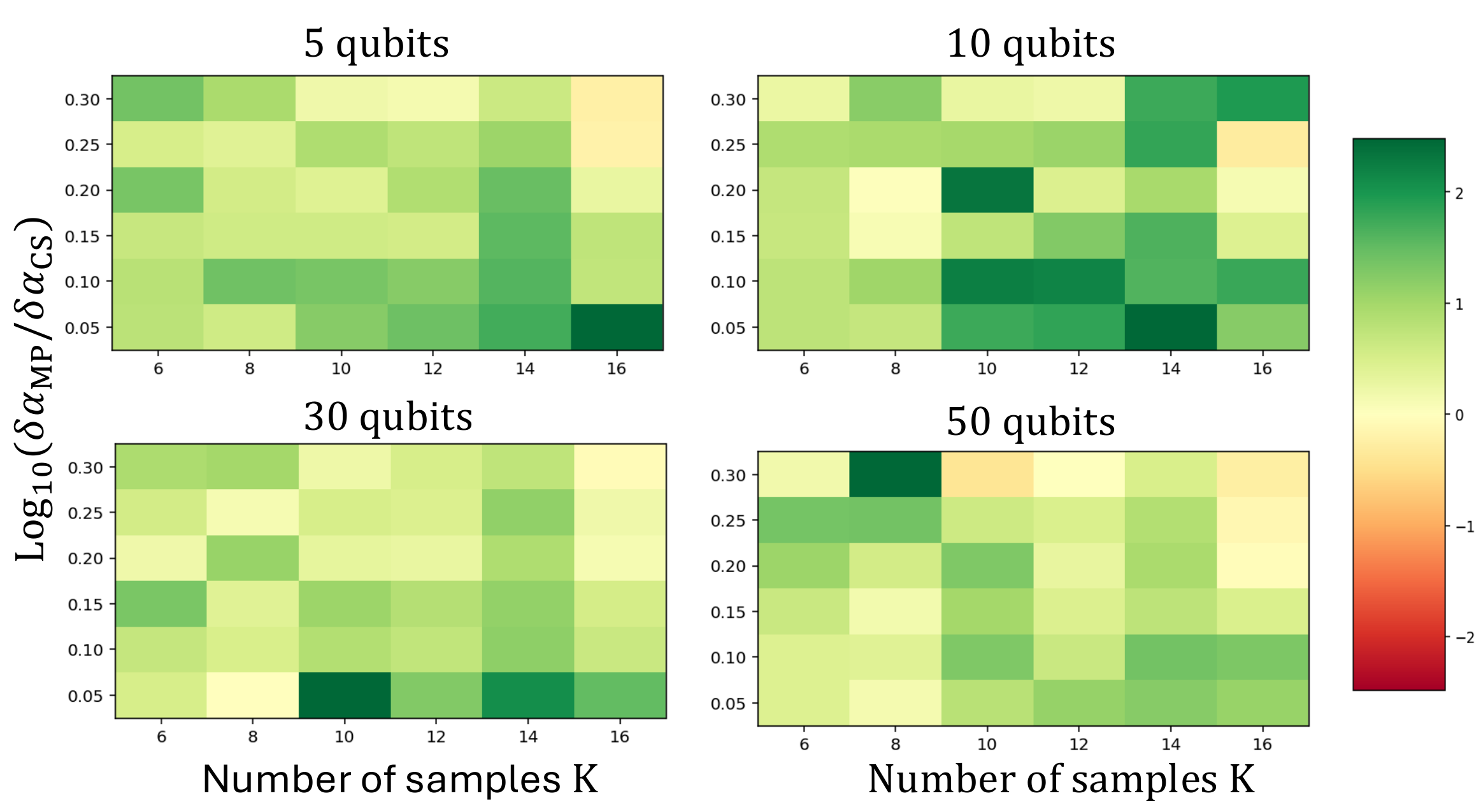}
\caption{{\bf Operating regime of the PSD-constrained reconstruction.} The color scale shows
\(\log_{10}(\delta\alpha_{\rm MP}/\delta\alpha_{\rm CS})\)
as a function of the number of retained samples \(K\) and the imperfection strength \(\gamma\), for \(n=5,10,30,\) and \(50\). Positive values indicate that the PSD-constrained reconstruction improves over the matrix-pencil unconstrained baseline. The results show that the method is most useful in the data-starved regime with moderate imperfections, while its advantage is reduced when the data are nearly ideal or when the imperfections become too strong.}
  \label{fig:phase_diagram}
\end{figure}

Figure~\ref{fig:phase_diagram} shows the quantity
\[
\log_{10}\!\left(
\frac{\delta\alpha_{\rm MP}}{\delta\alpha_{\rm CS}}
\right).
\]
Positive values indicate that the PSD-constrained reconstruction improves over the best unconstrained baseline, while negative values indicate that the best unconstrained baseline performs better.

The scan shows qualitatively similar behavior for \(n=5,10,30,\) and \(50\). The PSD-constrained reconstruction is often beneficial in the sparse-sampling regime with nonzero imperfections, but the improvement is not uniform across all $K$ and $\gamma$. For very weak imperfections, the advantage is limited because the measured signal is already close to a physical correlation function. For stronger imperfections, the improvement becomes less uniform, reflecting the expected limitation of the method under large noise or model mismatch. These results should be interpreted as an empirical map of the useful operating regime rather than as a universal performance guarantee.

\subsection{Robustness to hidden dephasing}
To address the robustness of the method to a more physically motivated noise mechanism, we also tested a dephasing model. In this test, the synthetic noiseless signal was replaced by
\begin{equation}
    G_\Gamma(t)=e^{-\Gamma t}\cos(n\alpha t),
\end{equation}
where \(\Gamma\) is a dephasing rate. Importantly, the reconstruction algorithm and the subsequent frequency estimation were not given \(\Gamma\); the same fitting and reconstruction pipeline used in the main text was applied. Thus, this test should be interpreted as a robustness check under model mismatch, rather than as a dephasing-calibrated estimator.
\begin{figure}[t]
  \centering
\includegraphics[width=0.85\columnwidth]{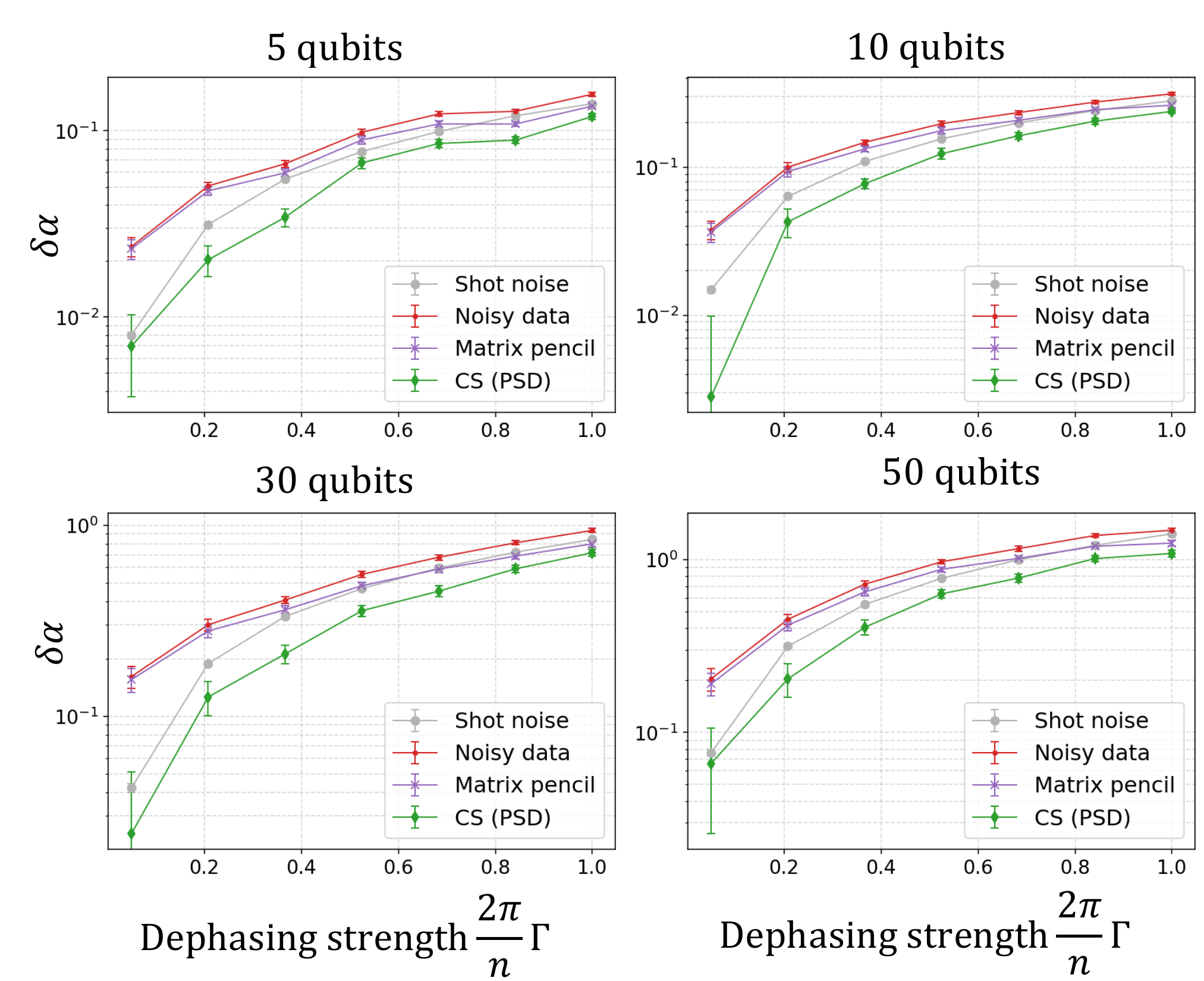}
\caption{{\bf Robustness to hidden dephasing.} The synthetic signal is generated as \(G_\Gamma(t)=e^{-\Gamma t}\cos(n\alpha t)\), while the estimators are not given the value of \(\Gamma\). The PSD-constrained reconstruction remains beneficial over the tested dephasing range, although the estimation error increases with \(\Gamma\) for all methods due to the increasing mismatch with the pure-cosine fitting model.}
  \label{fig:dephasing}
\end{figure}

Figure~\ref{fig:dephasing} shows the resulting estimation error as a function of \(\Gamma T\in[0.05,1]\). The dephasing range was chosen in units of the inverse observation window, so that the total damping over one oscillation period is varied systematically. As expected, all methods degrade as \(\Gamma\) increases, since the damped signal deviates increasingly from the pure-cosine model used for frequency extraction.  Nevertheless, over the tested range the PSD-constrained reconstruction yields lower estimation error than direct fitting of the noisy data and the matrix-pencil baseline. This supports the conclusion that the physicality constraint remains useful beyond the ideal undamped-cosine model, while also illustrating the expected limitation that strong model mismatch reduces the achievable accuracy.

\vfill

\end{document}